\documentclass[12pt]{article}
\usepackage{amsmath,amssymb,amsthm,braket,bm}
\usepackage{graphicx}
\usepackage{physics}
\usepackage{enumitem}
\usepackage[numbers]{natbib}
\usepackage{tikz}
\usetikzlibrary{quantikz2}
\usepackage{algorithm}
\usepackage{algpseudocode}
\usepackage{placeins}
\usepackage{booktabs}
\usepackage{tabularx}
\usepackage{array}
\usepackage{float}
\begin{document}

\title{Quantum Work Extraction via Conditional Spatial Displacements}

\author{
Necati Celik$^{*}$ 
}

\date{}

\maketitle

\begin{center}

Faculty of Engineering and Natural Sciences,\\
Department of Physics Engineering,\\
G\"um\"ushane University, 29100 G\"um\"ushane, Turkey\\
$^{*}$E-mail: necati.celik@gumushane.edu.tr

\vspace{0.4cm}

\end{center}

\begin{abstract}
We propose a protocol for extracting work from a coherent quantum 
battery state by exploiting measurement-assisted feedback mediated by a continuous-variable pointer. The scheme relies on the unitary operator $U = \exp(-i k t \, \hat{H} \otimes \hat{P}/\hbar)$, which generates entanglement between the battery's energy eigenstates and the position of an auxiliary pointer. A subsequent projective measurement of the pointer's position conditionally prepares the battery in a pure state from which work can be extracted via a feedback unitary. We analyze the protocol for a two-level quantum battery and a Gaussian pointer, computing the conditional states and the corresponding daemonic ergotropy. For the pure initial state considered, we find that the daemonic ergotropy equals the standard ergotropy for all interaction strengths, demonstrating that the measurement-assisted feedback recovers the full extractable work that would otherwise become inaccessible due to entanglement with the 
pointer when its degrees of freedom are traced out. The protocol thus provides a physically transparent realization of a quantum Maxwell demon, where the pointer acts as a quantum measurement ancilla whose position becomes correlated with the battery's energy. The scheme is amenable to experimental implementation in trapped-ion systems, and it contributes to the ongoing efforts to understand the role of quantum coherence and measurement in thermodynamics.
\end{abstract}

\newpage

\section{Introduction}

Extracting work from microscopic quantum systems lies at the heart of quantum thermodynamics. It has far-reaching consequences for how quantum devices—such as quantum batteries, heat engines, and refrigerators—are designed and operated \cite{Francica2017, Campaioli2024}. In this setting that we propose, ergotropy serves as a key figure of merit: it is the maximum amount of work that can be extracted from a quantum state through unitary operations \cite{Manzano2018}. Measurements and feedback can change how much work is accessible from a quantum system, because they provide information about its correlations with ancillary degrees of freedom \cite{Morrone2023, Elouard2017}. This idea goes back to Maxwell’s demon, a thought experiment in which an intelligent observer uses information about a system’s microscopic state to extract work, seemingly in violation of the second law of thermodynamics \cite{Kua2026, de2025friendly}.

Researchers have studied in considerable detail how measurement, information, and work extraction are connected in quantum systems \cite{Bernards2019}. One especially useful approach is the framework of daemonic ergotropy, which quantifies the maximum work that can be extracted from a quantum system after a measurement on a correlated ancillary system \cite{Simonov2022}. Recent theoretical and experimental studies have explored daemonic ergotropy in various settings, including Gaussian quantum states \cite{Kua2026}, open quantum batteries \cite{navid2025experimental}, and systems subjected to continuous monitoring \cite{Morrone2023}. Together, these studies have shown that measurement-assisted protocols can extract more work than the standard ergotropy allows, and that this enhancement is closely tied to measurement-induced purification \cite{Cenedese2026}. Recent work has also shown that entanglement in quantum batteries is not 
universally beneficial: depending on the protocol and the state preparation, 
entanglement can either enhance or degrade charging power, and the quantum 
state advantage of battery charging is not an entanglement monotone 
\cite{gyhm2024beneficial}.

In this work, we introduce a physically motivated protocol for measurement-assisted work extraction based on the unitary operator $U = \exp(-i k t \, \hat{H} \otimes \hat{P}/\hbar)$, where $\hat{H}$ is the Hamiltonian of the quantum battery, $\hat{P}$ is the momentum operator of an auxiliary pointer, and $k$ is a coupling constant with units of inverse momentum. This operator produces a conditional spatial displacement of the pointer, with the displacement determined by the battery’s energy. As a result, the battery’s energy eigenstates become entangled with the pointer’s position. Measuring the pointer’s position then reveals information about the battery’s energy without requiring a full projective measurement on the battery itself. This makes it possible to apply a conditional feedback operation that extracts work while preserving some of the quantum coherence. We study this protocol for a two-level quantum battery prepared in a pure initial state, together with a Gaussian pointer. We find that the daemonic ergotropy matches the standard ergotropy at all interaction strengths. In other words, the measurement-assisted feedback recovers the full work that would otherwise become inaccessible once the pointer is traced out and the battery is left entangled with it. The pointer therefore acts as a continuous-variable measurement ancilla, its position correlated with the battery’s energy, and it provides a physically transparent realization of Maxwell’s demon. Although the model does not yield any enhancement beyond the initial ergotropy, it clearly illustrates how quantum correlations and conditional feedback can preserve and recover work extraction when the battery is entangled with an unobserved pointer. The scheme is well suited to trapped-ion experiments and adds to the broader effort to understand the role of quantum coherence and measurement in thermodynamics.

\section{Model}
\label{sec:model}

We consider a quantum battery modeled as a two-level system with Hilbert
space $\mathcal{H}_S=\mathbb{C}^2$, spanned by the ground state
$\ket{g}$ and the excited state $\ket{e}$. The battery Hamiltonian is
chosen as
\begin{equation}
    \hat{H}
    =
    \epsilon \ket{e}\bra{e},
    \qquad
    \epsilon>0,
\label{eq:hamiltonian}
\end{equation}
where $\epsilon$ denotes the energy gap between the two levels. We choose
the zero of energy such that the ground-state energy is $E_g=0$ and the
excited-state energy is $E_e=\epsilon$. This choice of the energy
reference does not affect the dynamics or the ergotropy, since only energy
differences enter the work-extraction problem.

The battery is initially prepared in a coherent superposition of its
energy eigenstates,
\begin{equation}
    \ket{\psi_S}
    =
    \cos\frac{\theta}{2}\ket{g}
    +
    e^{i\phi}\sin\frac{\theta}{2}\ket{e},
\label{eq:initial_state}
\end{equation}
where $0\leq\theta\leq\pi$ and $0\leq\phi<2\pi$. The corresponding
density operator is
\begin{equation}
    \rho_S
    =
    \ket{\psi_S}\bra{\psi_S}
    =
    \begin{pmatrix}
        \cos^2\frac{\theta}{2}
        &
        \frac{1}{2}e^{-i\phi}\sin\theta
        \\[6pt]
        \frac{1}{2}e^{i\phi}\sin\theta
        &
        \sin^2\frac{\theta}{2}
    \end{pmatrix}.
\label{eq:initial_density}
\end{equation}
The off-diagonal elements in Eq.~(\ref{eq:initial_density}) quantify the
coherence between the ground and excited energy eigenstates.

Throughout this work, the time label \(t=0\) is omitted for initial
states and operators: \(\ket{\psi_S}\), \(\ket{\psi_P}\), and \(\rho_S\)
denote the initial battery state, pointer state, and battery density
operator, respectively. Time-evolved quantities are written with an
explicit argument, such as \(\ket{\Psi(t)}\) and \(\rho_S(t)\).

The battery is coupled to an auxiliary continuous-variable pointer with
Hilbert space
\begin{equation}
    \mathcal{H}_P=L^2(\mathbb{R}).
\end{equation}
The pointer is described by canonical position and momentum operators
$\hat{X}$ and $\hat{P}$ satisfying
$[\hat{X},\hat{P}]=i\hbar$.
We assume that the pointer is initially prepared in a normalized Gaussian
state with vanishing mean position and momentum,
$\langle \hat{X}\rangle_0=0,\qquad \langle \hat{P}\rangle_0=0$,
and position variance
$\langle \hat{X}^2\rangle_0=\sigma^2$.
Its wavefunction in the position representation is chosen to be
\begin{equation}
    \psi_P(x)
    =
    \braket{x}{\psi_P}
    =
    \frac{1}{(2\pi\sigma^2)^{1/4}}
    \exp\left(-\frac{x^2}{4\sigma^2}\right).
\label{eq:position_wavefunction}
\end{equation}
As known, the parameter $\sigma$ has dimensions of length and represents
the initial position uncertainty of the pointer.

Using the standard Fourier-transform convention as given below
\begin{equation}
    \tilde{\psi}_P(p)
    =
    \frac{1}{\sqrt{2\pi\hbar}}
    \int_{-\infty}^{\infty}
    dx\,
    e^{-ipx/\hbar}\psi_P(x),
\end{equation}
we optain the corresponding momentum-space wavefunction given as
\begin{equation}
    \tilde{\psi}_P(p)
    =
    \braket{p}{\psi_P}
    =
    \left(
        \frac{2\sigma^2}{\pi\hbar^2}
    \right)^{1/4}
    \exp\left(
        -\frac{\sigma^2p^2}{\hbar^2}
    \right).
\label{eq:momentum_wavefunction}
\end{equation}
This representation gives
$\langle \hat{P}\rangle_0=0,\qquad (\Delta P)^2=\frac{\hbar^2}{4\sigma^2}$,
and hence
$\Delta X\,\Delta P=\sigma\frac{\hbar}{2\sigma}=\frac{\hbar}{2}$.
Thus, the initial Gaussian pointer state is a minimum-uncertainty state.

The interaction between the battery and the pointer is described by the
unitary operator
\begin{equation}
    U
    =
    \exp\left(
        -\frac{i}{\hbar}
        \lambda\,\hat{H}\otimes\hat{P}
    \right),
\label{eq:unitary}
\end{equation}
where $\lambda$ is an interaction parameter. Since the exponent of a
unitary operator must be dimensionless, $\lambda$ must satisfy
$[\lambda] = \frac{\mathrm{length}}{\mathrm{energy}} = \frac{\mathrm{time}}{\mathrm{momentum}}$.
We write
$\lambda=kt$,
where $t$ is the interaction time and the coupling constant $k$ has
dimensions of inverse momentum,
$[k]=[\hat{P}]^{-1}$.
Consequently,
$[\lambda\hat{H}\hat{P}/\hbar]=1$,
as required for the exponential in Eq.~(\ref{eq:unitary}) to be
well-defined.

Because the Hamiltonian is diagonal in the $\{\ket{g},\ket{e}\}$ basis, we can express the unitary operator as the spectral decomposition form of
$\hat{H}$ as
\begin{equation}
    U
    =
    \ket{g}\bra{g}\otimes I_P
    +
    \ket{e}\bra{e}
    \otimes
    \exp\left(
        -\frac{i}{\hbar}
        \lambda\epsilon\hat{P}
    \right).
\label{eq:unitary_expanded}
\end{equation}
As can be seen, the second term in Eq.~(\ref{eq:unitary_expanded}) is a translation
operator in the pointer's position space. We therefore define the
conditional displacement as
\begin{equation}
    d
    =
    \lambda\epsilon
    =
    kt\epsilon.
\label{eq:displacement_parameter}
\end{equation}
Since $\lambda$ has dimensions of length per energy, $d$ has dimensions
of length. The translation operator then satisfies
\begin{equation}
    \exp\left(
        -\frac{i}{\hbar}d\hat{P}
    \right)
    \ket{\psi_P}
    =
    \ket{\psi_P+d}.
\label{eq:displacement}
\end{equation}
In the position representation, this relation becomes
\begin{align}
    \bra{x}
    \exp\left(
        -\frac{i}{\hbar}d\hat{P}
    \right)
    \ket{\psi_P}
    &=
    \psi_P(x-d).
\label{eq:wavefunction_displacement}
\end{align}
Thus, when the battery occupies $\ket{g}$, the pointer remains centered
at the origin, whereas when the battery occupies $\ket{e}$, the pointer
wavepacket is displaced by $d$.

Since it is useful, we introduced the dimensionless interaction parameter
\begin{equation}
    \chi
    =
    \frac{d}{\sigma}
    =
    \frac{\lambda\epsilon}{\sigma}
    =
    \frac{kt\epsilon}{\sigma}.
\label{eq:dimensionless_coupling}
\end{equation}
This parameter compares the conditional displacement of the pointer with
its intrinsic position uncertainty. The regimes
$\chi\ll1$, $\chi\sim1$, and $\chi\gg1$ correspond, respectively, to
weak, intermediate, and strong distinguishability of the two conditional
pointer states.

In order to see the action of the unitary operator on the joint state, we do the following
$\ket{\psi_S}\otimes\ket{\psi_P}$, the joint state after the
interaction is
\begin{align}
    \ket{\Psi(t)}
    &=
    U
    \left(
        \ket{\psi_S}
        \otimes
        \ket{\psi_P}
    \right)
    \nonumber\\
    &=
    \cos\frac{\theta}{2}
    \ket{g}\otimes\ket{\psi_P}
    +
    e^{i\phi}
    \sin\frac{\theta}{2}
    \ket{e}\otimes
    \ket{\psi_P+d}.
\label{eq:joint_state}
\end{align}
For nonzero $d$, the battery energy eigenstates become correlated with
different pointer positions. Except in special cases, the state in
Eq.~(\ref{eq:joint_state}) is therefore entangled.

The reduced state of the battery is obtained by tracing out the pointer as given below
\begin{equation}
    \rho_S(t)
    =
    \Tr_P
    \left[
        \ket{\Psi(t)}\bra{\Psi(t)}
    \right].
\label{eq:reduced_density}
\end{equation}
Expanding the joint density operator gives
\begin{align}
    \ket{\Psi(t)}\bra{\Psi(t)}
    &=
    \cos^2\frac{\theta}{2}
    \ket{g}\bra{g}
    \otimes
    \ket{\psi_P}\bra{\psi_P}
    \nonumber\\
    &\quad+
    \sin^2\frac{\theta}{2}
    \ket{e}\bra{e}
    \otimes
    \ket{\psi_P+d}\bra{\psi_P+d}
    \nonumber\\
    &\quad+
    \frac{1}{2}e^{-i\phi}\sin\theta\,
    \ket{g}\bra{e}
    \otimes
    \ket{\psi_P}
    \bra{\psi_P+d}
    \nonumber\\
    &\quad+
    \frac{1}{2}e^{i\phi}\sin\theta\,
    \ket{e}\bra{g}
    \otimes
    \ket{\psi_P+d}
    \bra{\psi_P}.
\label{eq:joint_density}
\end{align}
Taking the partial trace over the pointer gives
\begin{equation}
    \rho_S(t)
    =
    \begin{pmatrix}
        \cos^2\frac{\theta}{2}
        &
        \frac{1}{2}e^{-i\phi}\sin\theta\,
        \braket{\psi_P+d}{\psi_P}
        \\[8pt]
        \frac{1}{2}e^{i\phi}\sin\theta\,
        \braket{\psi_P}{\psi_P+d}
        &
        \sin^2\frac{\theta}{2}
    \end{pmatrix}.
\label{eq:reduced_matrix}
\end{equation}

The overlap between the two conditional pointer states determines the
remaining coherence of the battery. For the Gaussian state in
Eq.~(\ref{eq:position_wavefunction}), this overlap is
\begin{align}
    \braket{\psi_P}{\psi_P+d}
    &=
    \int_{-\infty}^{\infty}
    dx\,
    \psi_P^*(x)\psi_P(x-d)
    \nonumber\\
    &=
    \frac{1}{\sqrt{2\pi\sigma^2}}
    \int_{-\infty}^{\infty}
    dx\,
    \exp\left(
        -\frac{x^2}{4\sigma^2}
    \right)
    \exp\left(
        -\frac{(x-d)^2}{4\sigma^2}
    \right)
    \nonumber\\
    &=
    \exp\left(
        -\frac{d^2}{8\sigma^2}
    \right).
\label{eq:overlap_result}
\end{align}
As can be seen, the exponent is dimensionless because both $d$ and $\sigma$ have units of
length. In terms of the dimensionless interaction parameter $\chi$, we have written the 
overlap more compactly as
\begin{equation}
    \Gamma(\chi)
    =
    \exp\left(-\frac{\chi^2}{8}\right).
\label{eq:overlap_dimensionless}
\end{equation}

Consequently, the reduced state of the battery takes the form
\begin{equation}
    \rho_S(t)
    =
    \begin{pmatrix}
        \cos^2\frac{\theta}{2}
        &
        \frac{1}{2}e^{-i\phi}\sin\theta\,
        e^{-d^2/(8\sigma^2)}
        \\[8pt]
        \frac{1}{2}e^{i\phi}\sin\theta\,
        e^{-d^2/(8\sigma^2)}
        &
        \sin^2\frac{\theta}{2}
    \end{pmatrix}.
\label{eq:reduced_final}
\end{equation}
Equivalently, using Eq.~(\ref{eq:overlap_dimensionless}), the following can be obtained
\begin{equation}
    \rho_S(\chi)
    =
    \begin{pmatrix}
        \cos^2\frac{\theta}{2}
        &
        \frac{1}{2}e^{-i\phi}\sin\theta\,\Gamma(\chi)
        \\[8pt]
        \frac{1}{2}e^{i\phi}\sin\theta\,\Gamma(\chi)
        &
        \sin^2\frac{\theta}{2}
    \end{pmatrix}.
\label{eq:reduced_dimensionless}
\end{equation}

Equation~(\ref{eq:reduced_dimensionless}) shows that the interaction
preserves the populations of the battery's energy eigenstates while
reducing the magnitude of the energy-basis coherence. The coherence
factor is given in Equation~(\ref{eq:overlap_dimensionless}) 
which decreases monotonically from unity at $\chi=0$ to zero in the
limit $\chi\rightarrow\infty$. Therefore, the loss of coherence in the reduced
battery state originates from the entanglement generated between the
battery and the pointer. It is important to note that this is not an irreversible dissipative process for the joint system. The combined battery–pointer state stays pure and undergoes unitary evolution.

The reduced state is properly normalized, as can be seen from its trace,
$\Tr[\rho_S(t)] = \cos^2\frac{\theta}{2} + \sin^2\frac{\theta}{2} = 1$.
Its determinant is
$\det[\rho_S(t)] = \cos^2\frac{\theta}{2} \sin^2\frac{\theta}{2} - \frac{1}{4}\sin^2\theta\, \Gamma^2$,
which simplifies to
$\frac{1}{4}\sin^2\theta \left[ 1-\Gamma^2 \right] \geq 0$,
where $\Gamma = e^{-d^2/(8\sigma^2)}$.
The non-negativity of the determinant, together with unit trace and
Hermiticity, confirms that $\rho_S(t)$ is a valid density operator.
The purity of the reduced battery state is
$\Tr[\rho_S(t)^2] = 1 - \frac{1}{2}\sin^2\theta \left[ 1-\Gamma^2 \right]$.
Therefore, the purity decreases monotonically with increasing
$d/\sigma$, except for the limiting cases in which the initial state has
no coherence in the energy basis.

The limiting behavior further clarifies the role of the interaction.
For $d=0$, or equivalently $\chi=0$, the pointer states are identical and
the battery remains in its initial pure state as given below
\begin{equation}
    \rho_S
    =
    \ket{\psi_S}\bra{\psi_S}.
\end{equation}
In the opposite limit,
$\frac{d}{\sigma}\rightarrow\infty$,
the two conditional pointer states become orthogonal in the sense that
their overlap vanishes,
$\braket{\psi_P}{\psi_P+d} \rightarrow 0$.
The reduced battery state consequently approaches
\begin{equation}
    \rho_S(t)
    \longrightarrow
    \begin{pmatrix}
        \cos^2\frac{\theta}{2} & 0\\
        0 & \sin^2\frac{\theta}{2}
    \end{pmatrix}.
\label{eq:decohered_limit}
\end{equation}
Thus, in the strong-distinguishability limit, the reduced battery state
becomes an incoherent mixture of its two energy eigenstates.

The model therefore provides a continuous and analytically tractable
description of the transition from a coherent battery state to a
decohered reduced state through correlations with a continuous-variable
pointer. The dimensionless parameter $\chi=d/\sigma$ controls the degree
to which the two conditional pointer states can be distinguished. Small
values of $\chi$ correspond to strongly overlapping pointer states and
weak system--pointer distinguishability, whereas large values of $\chi$
correspond to nearly orthogonal pointer states and strong correlations
between the battery energy and the pointer position. This controlled
correlation provides the basis for the measurement-assisted work
extraction protocol considered in the following section.

\section{Conditional Work Extraction}
\label{sec:conditional_work}

\begin{figure}
    \centering
    \includegraphics[width=\linewidth]{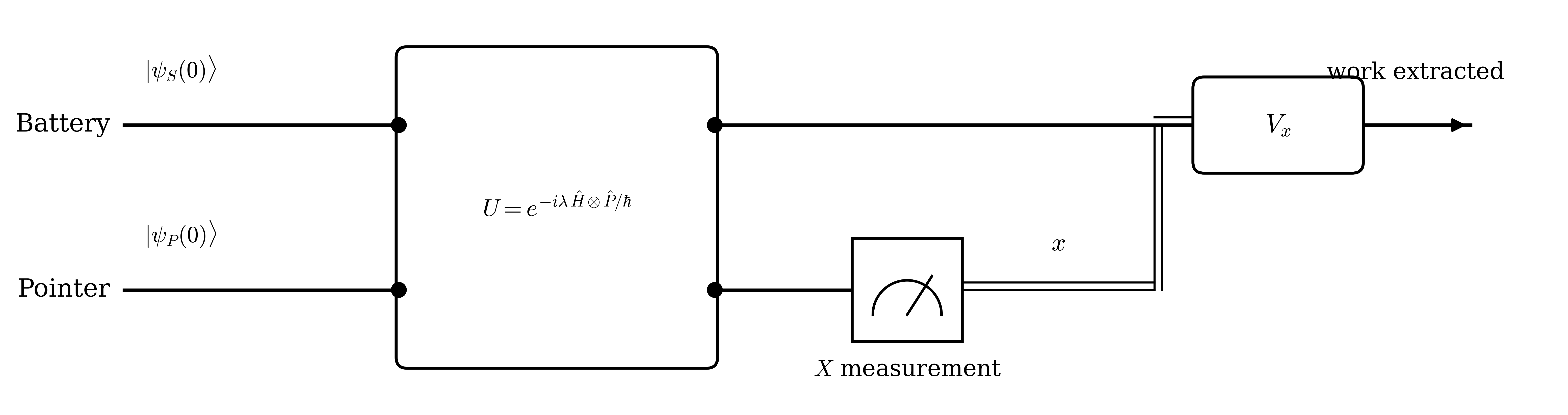}
    \caption{Schematic of the measurement-assisted work-extraction protocol.
    The battery and pointer are coupled by the unitary
    $U=e^{-i\lambda\hat{H}\otimes\hat{P}/\hbar}$, which correlates the
    battery's energy eigenstates with the pointer's position. A projective
    position measurement on the pointer yields outcome $x$, which is fed
    forward to implement the conditional unitary $V_x$ on the battery,
    extracting work.}
    \label{fig:protocol}
\end{figure}

Following the battery--pointer interaction described in
Sec.~\ref{sec:model}, we consider a measurement of the pointer position.
The purpose of this measurement is to obtain information about the
battery energy while retaining the conditional quantum state of the
battery. The full protocol is illustrated schematically in
Fig.~\ref{fig:protocol}.

An ideal measurement of the continuous pointer position is
described by the projection-valued measure
\begin{equation}
    \int_{-\infty}^{\infty} dx\,\ket{x}\bra{x}=I_P.
\label{eq:pointer_povm}
\end{equation}
Although position eigenstates are not normalizable in the usual
Hilbert-space sense, Eq.~(\ref{eq:pointer_povm}) is understood in the
standard generalized-eigenstate sense for a continuous-variable
measurement.

For a measurement outcome $x$, the unnormalized conditional state of the
battery is obtained by projecting the joint state onto $\ket{x}$:
\begin{equation}
    \ket{\widetilde{\psi}_{S|x}}
    =
    \bra{x}\Psi(t)\rangle.
\label{eq:unnormalized_conditional_state}
\end{equation}
Using the joint state obtained in Eq.~(\ref{eq:joint_state}), this becomes
\begin{equation}
    \ket{\widetilde{\psi}_{S|x}}
    =
    \cos\frac{\theta}{2}\,
    \psi_P(x)\ket{g}
    +
    e^{i\phi}\sin\frac{\theta}{2}\,
    \psi_P(x-d)\ket{e}.
\label{eq:unnormalized_conditional_explicit}
\end{equation}
The probability density associated with the measurement outcome $x$ is
the squared norm of the unnormalized state,
\begin{equation}
    p(x)
    =
    \braket{\widetilde{\psi}_{S|x}}
    {\widetilde{\psi}_{S|x}}.
\label{eq:probability_from_norm}
\end{equation}
Since $\langle g|e\rangle=0$, the cross terms vanish and therefore
\begin{equation}
    p(x)
    =
    \cos^2\frac{\theta}{2}
    |\psi_P(x)|^2
    +
    \sin^2\frac{\theta}{2}
    |\psi_P(x-d)|^2.
\label{eq:p_x_final}
\end{equation}
The probability density is normalized because
\begin{align}
    \int_{-\infty}^{\infty}dx\,p(x)
    &=
    \cos^2\frac{\theta}{2}
    \int_{-\infty}^{\infty}dx\,|\psi_P(x)|^2
    \nonumber\\
    &\quad+
    \sin^2\frac{\theta}{2}
    \int_{-\infty}^{\infty}dx\,|\psi_P(x-d)|^2
    \nonumber\\
    &=
    \cos^2\frac{\theta}{2}
    +
    \sin^2\frac{\theta}{2}
    =
    1.
\label{eq:p_x_normalization}
\end{align}

The normalized conditional state of the battery is consequently
\begin{equation}
    \ket{\psi_{S|x}}
    =
    \frac{
        \cos\frac{\theta}{2}\,
        \psi_P(x)\ket{g}
        +
        e^{i\phi}\sin\frac{\theta}{2}\,
        \psi_P(x-d)\ket{e}
    }
    {\sqrt{p(x)}}.
\label{eq:conditional_pure_state}
\end{equation}
The corresponding density operator is
\begin{equation}
    \rho_{S|x}
    =
    \ket{\psi_{S|x}}\bra{\psi_{S|x}},
\label{eq:conditional_density_pure}
\end{equation}
or explicitly,
\begin{equation}
    \rho_{S|x}
    =
    \frac{1}{p(x)}
    \begin{pmatrix}
        \displaystyle
        \cos^2\frac{\theta}{2}
        |\psi_P(x)|^2
        &
        \displaystyle
        \frac{1}{2}e^{-i\phi}\sin\theta\,
        \psi_P(x)\psi_P^*(x-d)
        \\[12pt]
        \displaystyle
        \frac{1}{2}e^{i\phi}\sin\theta\,
        \psi_P^*(x)\psi_P(x-d)
        &
        \displaystyle
        \sin^2\frac{\theta}{2}
        |\psi_P(x-d)|^2
    \end{pmatrix}.
\label{eq:conditional_matrix}
\end{equation}

An important property of Eq.~(\ref{eq:conditional_density_pure}) is that
the conditional state is pure for every measurement outcome $x$ for
which $p(x)>0$. This follows directly from the purity of the joint
battery--pointer state and the rank-one nature of the ideal position
measurement.In particular, $\Tr\left[\rho_{S|x}^2\right]=1$.
Thus, although the unconditional reduced battery state becomes mixed
because of its entanglement with the pointer, conditioning on a specific
pointer outcome restores a pure state of the battery.

The excited-state population of the conditional battery state is given as
\begin{equation}
    \rho_{ee}(x)
    =
    \bra{e}\rho_{S|x}\ket{e}
    =
    \sin^2\frac{\theta}{2}
    \frac{|\psi_P(x-d)|^2}{p(x)}.
\label{eq:conditional_population}
\end{equation}
Similarly, the conditional coherence is given as
\begin{equation}
    \rho_{eg}(x)
    =
    \frac{1}{2}e^{i\phi}\sin\theta\,
    \frac{
        \psi_P^*(x)\psi_P(x-d)
    }
    {p(x)}.
\label{eq:conditional_coherence}
\end{equation}
The conditional coherence is strongest in the region where the two pointer wavefunctions overlap. This coherence, however, should not be taken as evidence of enhanced work extraction. Since the conditional states in this model are pure, their ergotropy follows entirely from their mean energy.

For the sake of completeness, consider a general two-level density matrix given as
\begin{equation}
    \rho=
    \begin{pmatrix}
        \rho_{gg} & \rho_{ge}\\
        \rho_{eg} & \rho_{ee}
    \end{pmatrix},
\end{equation}
with Hamiltonian given in Equation \ref{eq:hamiltonian}.
The mean energy is therefore given as
\begin{equation}
    E(\rho)
    =
    \Tr(\rho\hat{H})
    =
    \epsilon\rho_{ee}.
\label{eq:mean_energy_general}
\end{equation}
The eigenvalues of $\rho$ are
\begin{equation}
    r_{\pm}
    =
    \frac{1}{2}
    \left[
        1
        \pm
        \sqrt{
            (2\rho_{ee}-1)^2
            +
            4|\rho_{eg}|^2
        }
    \right].
\label{eq:rho_eigenvalues}
\end{equation}
For a two-level system, the passive state is obtained by assigning the
larger eigenvalue to the ground state and the smaller eigenvalue to the
excited state. Consequently, the ergotropy is
\begin{equation}
    \mathcal{E}(\rho)
    =
    \epsilon
    \left[
        \rho_{ee}
        -
        r_-
    \right],
\label{eq:general_ergotropy}
\end{equation}
or equivalently
\begin{equation}
    \mathcal{E}(\rho)
    =
    \frac{\epsilon}{2}
    \left[
        2\rho_{ee}-1
        +
        \sqrt{
            (2\rho_{ee}-1)^2
            +
            4|\rho_{eg}|^2
        }
    \right].
\label{eq:ergotropy_formula}
\end{equation}

For the conditional state in Eq.~(\ref{eq:conditional_density_pure}),
the purity condition implies that its eigenvalues are $1$ and $0$.
Therefore, Eq.~(\ref{eq:ergotropy_formula}) simplifies considerably:
\begin{equation}
    \mathcal{E}(\rho_{S|x})
    =
    \epsilon\rho_{ee}(x).
\label{eq:conditional_ergotropy}
\end{equation}
Substituting Eq.~(\ref{eq:conditional_population}) gives
\begin{equation}
    \mathcal{E}(\rho_{S|x})
    =
    \epsilon
    \sin^2\frac{\theta}{2}
    \frac{
        |\psi_P(x-d)|^2
    }
    {p(x)}.
\label{eq:conditional_ergotropy_explicit}
\end{equation}

For the chosen projective measurement of the pointer position, the 
average work that can be extracted when the measurement outcome 
\(x\) is available for conditional feedback is
\begin{equation}
    \mathcal{W}_D^{(X)}
    =
    \int_{-\infty}^{\infty}
    dx\,
    p(x)\,
    \mathcal{E}(\rho_{S|x}).
\label{eq:measurement_assisted_work}
\end{equation}
Using Eq.~(\ref{eq:conditional_ergotropy_explicit}), we obtain
\begin{align}
    \mathcal{W}_D^{(X)}
    &=
    \epsilon
    \sin^2\frac{\theta}{2}
    \int_{-\infty}^{\infty}
    dx\,
    |\psi_P(x-d)|^2
    \nonumber\\
    &=
    \epsilon
    \sin^2\frac{\theta}{2},
\label{eq:daemonic_ergotropy_result}
\end{align}
where normalization of the displaced pointer state has been used:
\begin{equation}
    \int_{-\infty}^{\infty}
    dx\,
    |\psi_P(x-d)|^2
    =
    1.
\end{equation}
Therefore, the measurement-assisted extractable work for the chosen 
position measurement has the exact value
$\mathcal{W}_D^{(X)} = \epsilon\sin^2\frac{\theta}{2}$
for every value of the displacement $d$.
This quantity fits within the daemonic-ergotropy framework
\cite{Cheong2023, Pushpan2026}. In fact, for the pure joint
state considered here, it coincides with the globally optimised
daemonic ergotropy. For any pointer measurement, the ergotropy
of each conditional battery state is bounded by its mean energy.
Since the interaction commutes with the battery Hamiltonian, the
average mean energy is fixed at the initial value. A rank-one
measurement, such as the position PVM, produces pure conditional
states, for which the ergotropy equals the mean energy, thereby
saturating the bound. Hence Eq.~(\ref{eq:daemonic_ergotropy_result}) is already globally optimal, and any fully resolved rank-one
pointer measurement would yield the same average work. The
superscript \(X\) indicates that this quantity is associated
with the measurement of the pointer's position.

This result has an important consequence for the present protocol. The
interaction
$U = \exp\left( -\frac{i}{\hbar} \lambda\hat{H}\otimes\hat{P} \right)$
commutes with the battery Hamiltonian as
\begin{equation}
    [U,\hat{H}\otimes I_P]=0.
\label{eq:energy_commutation}
\end{equation}
Consequently, the interaction does not change the populations of the
battery energy eigenstates. In particular,
\begin{equation}
    \Tr[\rho_S(t)\hat{H}]
    =
    \Tr[\rho_S\hat{H}]
    =
    \epsilon\sin^2\frac{\theta}{2}.
\label{eq:energy_conservation}
\end{equation}
The pointer interaction therefore redistributes the accessibility of the
battery energy by transferring coherence into system--pointer
correlations, but it does not supply additional energy to the battery.

The same result can be understood directly from the conditional
feedback protocol. For a given outcome $x$, the conditional state
$\ket{\psi_{S|x}}$ is pure. Hence an outcome-dependent unitary
$V_x$ can rotate this state to the ground state,
\begin{equation}
    V_x\ket{\psi_{S|x}}
    =
    \ket{g},
\label{eq:optimal_feedback}
\end{equation}
thereby extracting all of its mean energy. The extracted work associated
with outcome $x$ is therefore given as
\begin{equation}
    W_x
    =
    \Tr[\rho_{S|x}\hat{H}]
    =
    \epsilon\rho_{ee}(x).
\label{eq:conditional_extracted_work}
\end{equation}
Averaging over all measurement outcomes gives
\begin{equation}
    \langle W_x\rangle
    =
    \int_{-\infty}^{\infty}
    dx\,p(x)W_x
    =
    \epsilon\sin^2\frac{\theta}{2}.
\label{eq:average_extracted_work}
\end{equation}
Thus, in the idealized protocol considered here, the measurement and
conditional feedback recover the complete ergotropy of the initial pure
battery state.

The result in Eq.~(\ref{eq:daemonic_ergotropy_result}) should not be read as an
enhancement of the total extractable work beyond the initial ergotropy.
Rather, the pointer measurement makes conditional work extraction possible
once the battery has become correlated with the pointer. As $d/\sigma$
increases, the unconditional reduced battery state loses coherence and its
ergotropy drops accordingly. Knowing the pointer outcome gives access to the
correlations produced by the interaction, allowing the corresponding
conditional pure state to be used optimally.

The limiting cases provide useful checks of the result. For
$d\rightarrow0$, the two pointer states coincide,
$\ket{\psi_P+d} \rightarrow \ket{\psi_P}$,
and the conditional battery state becomes independent of $x$:
$\rho_{S|x} \rightarrow \rho_S$.
Consequently,
\begin{equation}
\mathcal{W}_D^{(X)}(d\rightarrow0)
=
\mathcal{E}(\rho_S)
=
\epsilon\sin^2\frac{\theta}{2}.
\label{eq:wd_d0}
\end{equation}

In the opposite limit, $d/\sigma\rightarrow\infty$, the two pointer
wavepackets become effectively distinguishable:
\begin{equation}
    \braket{\psi_P}{\psi_P+d}
    \rightarrow0.
\end{equation}
For a given measurement outcome, the conditional state approaches one of the
energy eigenstates, apart from an exponentially small region where the two
pointer distributions still overlap. The conditional states therefore become
nearly diagonal, and the measurement outcome provides almost complete
information about the battery energy. The average extractable work remains
\begin{equation}
\mathcal{W}_D^{(X)}(d/\sigma\rightarrow\infty)
=
\epsilon\sin^2\frac{\theta}{2}.
\label{eq:wd_dinfty}
\end{equation}

Taken together, Eqs.~\eqref{eq:wd_d0} and \eqref{eq:wd_dinfty} and the exact
result in Eq.~\eqref{eq:daemonic_ergotropy_result} show that the
measurement-assisted extractable work \(\mathcal{W}_D^{(X)}\) does not depend
on the pointer displacement in this idealized model. What the displacement
does control is the coherence left in the unconditional battery state and how
well the conditional pointer states can be distinguished. The main effect of
the measurement protocol is therefore not to create extra energy or to boost
the initial ergotropy, but to recover---through conditional feedback---the
extractable work that would otherwise be lost once the pointer degrees of
freedom are ignored.

\section{Comparison and Role of Quantum Coherence}

To assess the performance of our protocol, we compare the daemonic ergotropy \(\mathcal{W}_D\) with the standard ergotropy \(\mathcal{E}(\rho_S)\) of the initial battery state, and with the work extractable via a projective energy measurement on the battery. This comparison illuminates the distinct roles of measurement back-action and quantum coherence in the present model.

The standard ergotropy of the initial coherent state can be obtained from the general formula in Eq.~(\ref{eq:ergotropy_formula}). Recall that the initial density matrix is given by Eq.~(\ref{eq:initial_density}), so the excited-state population and coherence are
\begin{equation}
\rho_{ee} = \sin^2\frac{\theta}{2}, \qquad \rho_{eg} = \frac{1}{2} e^{i\phi}\sin\theta.
\label{eq:initial_pop_coherence}
\end{equation}
Substituting these into Eq.~(\ref{eq:ergotropy_formula}) gives
\begin{align}
\mathcal{E}(\rho_S)
&= \epsilon \left( \sin^2\frac{\theta}{2} - \frac{1}{2} + \frac{1}{2}\sqrt{(2\sin^2\frac{\theta}{2} - 1)^2 + 4|\rho_{eg}|^2} \right) \nonumber \\
&= \epsilon \left( \sin^2\frac{\theta}{2} - \frac{1}{2} + \frac{1}{2}\sqrt{(-\cos\theta)^2 + \sin^2\theta} \right) \nonumber \\
&= \epsilon \left( \sin^2\frac{\theta}{2} - \frac{1}{2} + \frac{1}{2} \right) \nonumber \\
&= \epsilon \sin^2\frac{\theta}{2}.
\label{eq:standard_ergotropy}
\end{align}
This is the maximum work that can be extracted from the initial state without any measurement or feedback. It depends only on the excited-state population and is independent of the phase \(\phi\). This result has a clear physical interpretation: for the pure initial state considered here, the ergotropy equals the entire mean energy, so there is no additional work beyond this amount that can be attributed separately to coherence. The optimal unitary operation that extracts this work is a rotation that maps the initial pure state to the ground state: \(V\ket{\psi_S} = \ket{g}\).
The extracted work is then the decrease in the battery's internal energy:
\begin{equation}
W = \Tr[\rho_S\hat{H}] - \Tr[\ket{g}\bra{g}\hat{H}]
  = \epsilon\sin^2\frac{\theta}{2} - 0
  = \epsilon\sin^2\frac{\theta}{2}.
\end{equation}
This confirms that all the energy stored in the excited-state population 
can be extracted, while the coherence does not contribute additional 
energy beyond the population already present.

The unconditional reduced state of the battery after the interaction with
the pointer is given in Eq.~(\ref{eq:reduced_dimensionless}):
\begin{equation}
    \rho_S(d)
    =
    \begin{pmatrix}
        \cos^2\frac{\theta}{2}
        &
        \frac{1}{2}e^{-i\phi}\sin\theta\,\Gamma(d)
        \\[8pt]
        \frac{1}{2}e^{i\phi}\sin\theta\,\Gamma(d)
        &
        \sin^2\frac{\theta}{2}
    \end{pmatrix},
    \qquad
    \Gamma(d)=e^{-d^2/(8\sigma^2)}.
\label{eq:reduced_state_for_ergotropy}
\end{equation}
The populations are unchanged, but the coherence is suppressed by the
factor \(\Gamma(d)\). The ergotropy of this mixed state is obtained from
the general formula Eq.~(\ref{eq:ergotropy_formula}). With
\(\rho_{ee}=\sin^2(\theta/2)\) and
\(|\rho_{eg}|=\frac12\sin\theta\,\Gamma(d)\), we find
\begin{align}
    \mathcal E[\rho_S(d)]
    &=
    \frac{\epsilon}{2}
    \Bigg[
        2\sin^2\frac{\theta}{2}-1
        +
        \sqrt{
            \left(2\sin^2\frac{\theta}{2}-1\right)^2
            +
            \sin^2\theta\,\Gamma(d)^2
        }
    \Bigg] \nonumber\\
    &=
    \frac{\epsilon}{2}
    \left[
        -\cos\theta
        +
        \sqrt{
            \cos^2\theta
            +
            e^{-d^2/(4\sigma^2)}\sin^2\theta
        }
    \right],
\label{eq:post_interaction_ergotropy}
\end{align}
where we used \(\Gamma(d)^2=e^{-d^2/(4\sigma^2)}\). This expression
is the explicit form of the ergotropy of the reduced battery state after
the system–pointer interaction but before any measurement or feedback.

For \(d=0\) (\(\Gamma=1\)), Eq.~(\ref{eq:post_interaction_ergotropy})
reduces to
\begin{equation}
\mathcal E[\rho_S]
=
\frac{\epsilon}{2}\left[-\cos\theta + 1\right]
=
\epsilon\sin^2\frac{\theta}{2},
\end{equation}
which is the initial ergotropy. For \(d\to\infty\) (\(\Gamma\to0\)), it
becomes
\begin{equation}
\mathcal E[\rho_S(\infty)]
=
\frac{\epsilon}{2}\left[-\cos\theta + |\cos\theta|\right]
=
\begin{cases}
0, & \theta\le \pi/2,\\[4pt]
-\epsilon\cos\theta, & \theta\ge \pi/2,
\end{cases}
\end{equation}
which equals \(0\) for \(\theta\le\pi/2\) (the state is passive) and 
\(-\epsilon\cos\theta = \epsilon(\sin^2\frac{\theta}{2} - \cos^2\frac{\theta}{2})\) 
for \(\theta\ge\pi/2\), corresponding to the work extractable by swapping 
the populations of the fully decohered mixture.In the general case, the derivative of
Eq.~(\ref{eq:post_interaction_ergotropy}) with respect to
\(\chi=d/\sigma\) is
\begin{equation}
\frac{d}{d\chi}\mathcal E[\rho_S(\chi)]
=
-\frac{\epsilon\,\chi\,\sin^2\theta}{8\sqrt{\cos^2\theta+e^{-\chi^2/4}\sin^2\theta}}
\,e^{-\chi^2/4}\le 0,
\end{equation}
with equality only at \(\theta=0,\pi\) or \(\chi=0\). Thus, for
\(0<\theta<\pi\), the ergotropy of the unconditioned state decreases
monotonically with the interaction strength, as expected from the loss
of coherence.

Comparing Eq.~(\ref{eq:post_interaction_ergotropy}) with the 
measurement-assisted extractable work \(\mathcal{W}_D^{(X)}\) 
obtained in Eq.~(\ref{eq:daemonic_ergotropy_result}) and the initial
ergotropy in Eq.~(\ref{eq:standard_ergotropy}), we obtain the central
inequality
\begin{equation}
\mathcal E[\rho_S(d)]
\;\le\;
\mathcal{W}_D^{(X)}
=
\mathcal E[\rho_S]
=
\epsilon\sin^2\frac{\theta}{2}.
\label{eq:central_inequality}
\end{equation}

\begin{figure}[htbp]
    \centering
    \includegraphics[width=0.75\linewidth]{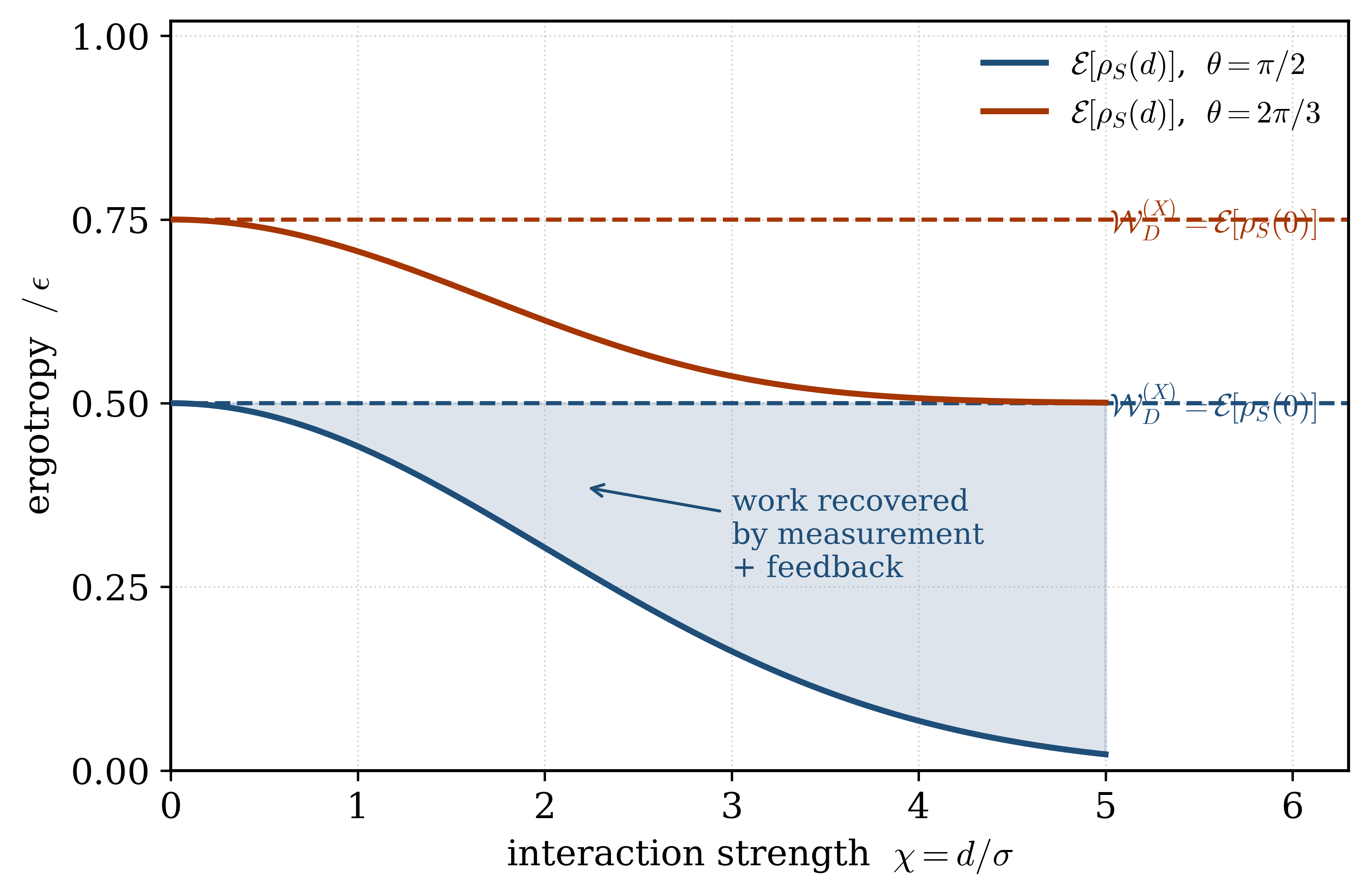}
    \caption{Ergotropy (in units of $\epsilon$) as a function of the
    dimensionless interaction strength $\chi=d/\sigma$. Solid curves:
    ergotropy $\mathcal{E}[\rho_S(d)]$ of the unconditional reduced
    battery state after the system--pointer interaction, shown for
    $\theta=\pi/2$ (blue) and $\theta=2\pi/3$ (orange). Dashed horizontal
    lines: the constant daemonic/initial ergotropy
    $\mathcal{W}_D^{(X)}=\mathcal{E}[\rho_S]=\epsilon\sin^2(\theta/2)$.
    The shaded region (blue) is the work that would be inaccessible if the
    pointer were traced out, but is recovered by the pointer measurement
    together with conditional feedback.}
    \label{fig:ergotropy}
\end{figure}

For any finite \(d>0\) and \(0<\theta<\pi\), the inequality is strict.
This demonstrates quantitatively that the unconditional reduced state
loses extractable work, but the measurement-assisted feedback protocol
recovers the full initial ergotropy. The loss is precisely the amount
of work that becomes inaccessible when the pointer degrees of freedom
are ignored, and it is exactly compensated by using the measurement
outcome to condition the feedback operation. This inequality is
illustrated in Fig.~\ref{fig:ergotropy}.

For comparison, we consider the case where a projective energy measurement is performed directly on the battery. In this case, the measurement projects the battery onto one of its energy eigenstates:
\begin{equation}
\ket{\psi_S} \to 
\begin{cases}
\ket{g} & \text{with probability } \cos^2\frac{\theta}{2}, \\
\ket{e} & \text{with probability } \sin^2\frac{\theta}{2}.
\end{cases}
\label{eq:projective_measurement}
\end{equation}
After the measurement, the state of the battery is either \(\ket{g}\) with probability \(\cos^2(\theta/2)\) or \(\ket{e}\) with probability \(\sin^2(\theta/2)\). The ergotropy of the ground state is zero, since no unitary operation can lower its energy below the ground state:
\begin{equation}
\mathcal{E}(\ket{g}\bra{g}) = 0.
\label{eq:ground_state_ergotropy}
\end{equation}
In contrast, the ergotropy of the excited state is \(\epsilon\), because a unitary operation can rotate the excited state to the ground state and extract the full energy:
\begin{equation}
\mathcal{E}(\ket{e}\bra{e}) = \epsilon.
\label{eq:excited_state_ergotropy}
\end{equation}
Therefore, the average work extractable after a projective energy measurement, followed by optimal work extraction from the resulting eigenstate, is
\begin{equation}
\mathcal{W}_{\text{proj}} = \cos^2\frac{\theta}{2} \cdot 0 + \sin^2\frac{\theta}{2} \cdot \epsilon = \epsilon \sin^2\frac{\theta}{2} = \mathcal{E}(\rho_S).
\label{eq:projective_work_corrected}
\end{equation}
Thus, a projective energy measurement followed by optimal work extraction yields the same amount of work as the standard ergotropy of the initial state. The measurement itself does not reduce the extractable work from the population; it simply collapses the coherence. However, it does not allow for any enhancement beyond the standard ergotropy, because it destroys all coherence.

As shown previously, the measurement-assisted extractable work for our protocol is
\begin{equation}
\mathcal{W}_D^{(X)} = \epsilon\sin^2\frac{\theta}{2} = \mathcal{E}(\rho_S),
\label{eq:wd_exact_comparison}
\end{equation}
for every value of the displacement \(d\). Therefore, the pointer measurement protocol yields the same average extractable work as both the standard ergotropy and the projective energy measurement. The equality
\begin{equation}
\mathcal{W}_D^{(X)} = \mathcal{W}_{\text{proj}} = \mathcal{E}(\rho_S)
\label{eq:all_equal}
\end{equation}
holds for all values of the interaction strength.

This finding is consistent with the broader observation that quantum 
correlations do not automatically translate into a thermodynamic advantage. 
In particular, Gyhm and Fischer have shown that highly entangled states can 
be detrimental for quantum battery charging power, and that the quantum state 
advantage is not an entanglement monotone \cite{gyhm2024beneficial}. In the present 
protocol, the equality $\mathcal{W}_D^{(X)} = \mathcal{E}(\rho_S)$ likewise 
shows that the entanglement generated between the battery and the pointer 
does not enhance the extractable work beyond the initial ergotropy; it only 
preserves the work that would otherwise become inaccessible.

Despite this equality, the two measurement protocols differ 
fundamentally in their treatment of quantum coherence and their 
experimental requirements. In the projective energy measurement, 
the coherence of the initial state is completely destroyed, and 
the resulting conditional states are energy eigenstates with no 
coherence. In the pointer measurement protocol, the conditional 
states retain coherence for finite \(d\). This coherence is present 
in the off-diagonal elements of \(\rho_{S|x}\) and can, in principle, 
be exploited for subsequent operations beyond simple work extraction. 
However, because the initial state is pure and the interaction 
commutes with the battery Hamiltonian, the average extractable work 
remains equal to the initial value. 

Thus, the pointer protocol is not superior to the direct energy 
measurement in terms of the amount of work extracted. Its advantage 
lies instead in the preservation of quantum coherence, which may be 
valuable in contexts where the conditional state must be manipulated 
further after the measurement, or where coherence itself serves as 
a resource for subsequent quantum information processing tasks 
\cite{Campaioli2024}. In contrast, the direct energy measurement 
irreversibly destroys all coherence, yielding conditional states 
that are strictly diagonal in the energy basis.

\begin{table}[htbp]
\centering
\caption{Comparison of the two measurement protocols. Both yield the
same average extractable work, but the pointer position measurement
preserves conditional coherence, whereas the direct energy measurement
destroys it.}
\label{tab:comparison}
\renewcommand{\arraystretch}{1.2}
\begin{tabularx}{\linewidth}{@{} >{\raggedright\arraybackslash}p{0.28\linewidth}
                                 >{\centering\arraybackslash}p{0.28\linewidth}
                                 >{\raggedright\arraybackslash}X @{}}
\toprule
\textbf{Protocol} & \textbf{Extractable work} & \textbf{Conditional state} \\
\midrule
Direct energy measurement
  & \(\epsilon\sin^2(\theta/2)\)
  & \(\ket{g}\) or \(\ket{e}\) (no coherence) \\
Pointer position measurement
  & \(\mathcal{W}_D^{(X)} = \epsilon\sin^2(\theta/2)\)
  & Coherent superposition (coherence preserved) \\
\bottomrule
\end{tabularx}
\end{table}

The limiting cases provide useful checks of the general result. In the limit \(d \to 0\) (no interaction), the two conditional pointer states coincide, and the measurement provides no information about the battery. The conditional states become
\begin{equation}
\rho_{S|x} \to \rho_S,
\label{eq:conditional_d0}
\end{equation}
and the daemonic ergotropy reduces to the standard ergotropy:
\begin{equation}
\mathcal{W}_D(d \to 0) = \mathcal{E}(\rho_S).
\label{eq:wd_d0_1}
\end{equation}
In this limit, no information is gained, and the protocol reduces to standard work extraction from the initial state. 
In the opposite limit \(d \to \infty\) (strong interaction), the two 
conditional pointer states become orthogonal, and the measurement 
provides complete which-energy information. However, because the 
conditional state remains pure for every finite \(d\), it cannot become 
a generic diagonal mixed state. Instead, for almost every measurement 
outcome, the conditional state approaches one of the two energy 
eigenstates, depending on which pointer packet the outcome \(x\) 
originates from. For a fixed finite \(x\), we have 
\(\psi_P(x-d) \to 0\), and therefore
$\rho_{S|x} \longrightarrow \ket{g}\bra{g}$.
For outcomes near the displaced packet, i.e., \(x-d\) finite while 
\(x \to \infty\), we have \(\psi_P(x) \to 0\), and therefore
$\rho_{S|x} \longrightarrow \ket{e}\bra{e}$.
In the limit \(d/\sigma\rightarrow\infty\), the probability weight of 
the region in which both conditional wavefunctions contribute 
significantly vanishes. Thus, in the strong-interaction limit, the 
measurement perfectly distinguishes the two energy eigenstates, and the 
conditional states are the corresponding eigenstates, with no remaining 
coherence. This is consistent with the purity condition 
\(\Tr(\rho_{S|x}^2)=1\). The daemonic ergotropy in this limit remains
$\mathcal{W}_D(d \to \infty) = \epsilon\sin^2\frac{\theta}{2} = \mathcal{E}(\rho_S)$.
Thus, although the strong-interaction limit provides perfect which-energy information, the average extractable work is the same as in the no-interaction limit. The difference lies in the nature of the conditional states: in the strong-interaction limit, they are diagonal energy eigenstates, whereas in the weak-interaction limit, they are coherent superpositions.

For intermediate values of \(d\), the daemonic ergotropy interpolates smoothly between these two limits while maintaining the constant value \(\mathcal{E}(\rho_S)\). The conditional states retain partial coherence, and the measurement provides partial information about the battery energy. However, the average extractable work remains unchanged. This is a consequence of the fact that the conditional states are pure for every measurement outcome, and their ergotropy is determined solely by their mean energy, which averages to the initial mean energy.

The distinction between the pointer measurement and the projective energy measurement becomes particularly apparent when considering the role of quantum coherence for subsequent protocols. In the projective energy measurement, the coherence is irreversibly destroyed, and the conditional states are energy eigenstates with no coherence. In the pointer measurement, the conditional states retain coherence for finite \(d\), even though the total average extractable work remains the same. This coherence preservation may be beneficial in contexts where the conditional state must be manipulated further after the measurement, or where the coherence itself is a resource for subsequent quantum information processing tasks.

The second law of thermodynamics imposes fundamental bounds on the 
extractable work in the presence of feedback \cite{Sagawa2008}. In particular, the work extracted cannot exceed the decrease 
in free energy of the joint system, including the cost of the measurement 
and feedback operations. For our protocol, the measurement and feedback 
operations require work to be performed on the pointer and to implement 
the conditional unitary. The net work extracted is
\begin{equation}
\mathcal{W}_{\text{net}} = \mathcal{W}_D^{(X)} - \mathcal{W}_{\text{meas,cost}} - \mathcal{W}_{\text{fb,cost}},
\label{eq:net_work_corrected}
\end{equation}
where \(\mathcal{W}_{\text{meas,cost}}\) is the work cost of the 
position measurement and \(\mathcal{W}_{\text{fb,cost}}\) is the work 
cost of the feedback operation. These costs must be accounted for to 
ensure that the second law is not violated.

Equation~(\ref{eq:net_work_corrected}) accounts only for the
measurement and feedback stages of the protocol. A complete cyclic
accounting would also include the work supplied by the state-dependent
force, the work required to restore the pointer to its initial state,
and the work required to reset the measurement record. In the
trapped-ion realization, for instance, the conditional displacement
raises the energy of the motional mode, so this pointer energy cost
must be supplied by the external fields. Reusing the pointer and
erasing the measurement record likewise require additional work.
These terms are omitted in the idealized expression above, but they
must be included in any realistic thermodynamic balance \cite{de2025friendly}.

The work cost of a measurement depends on the specific experimental 
implementation and the nature of the detector. For continuous-variable 
measurements such as homodyne detection, the cost is associated with 
the energy required to couple the pointer to the detector and to 
process the measurement outcome. Similarly, the feedback operation 
requires work to implement the conditional unitary, which in trapped-ion 
systems can be realized by laser pulses whose timing and phase are 
controlled by the measurement outcome. The energy cost of these pulses 
must be included in \(\mathcal{W}_{\text{fb,cost}}\).

Recent studies have established tight bounds on the work costs of 
feedback control and have demonstrated that the 
second law is not violated when these costs are properly accounted for 
\cite{minagawa2025universal}. A detailed thermodynamic analysis of these costs is 
beyond the scope of this work, but our results provide the theoretical 
foundation for such an analysis in future studies. In the idealized 
limit where the measurement and feedback costs are negligible, the 
net work extracted equals the initial ergotropy. In any realistic 
implementation, the net work will be reduced by these costs, and a 
careful experimental design is required to ensure that
\[
\mathcal{W}_{\text{meas,cost}} + \mathcal{W}_{\text{fb,cost}} < 
\mathcal{W}_D^{(X)}
\]
so that positive net work extraction is achieved.

Finally, we emphasize that what this protocol mainly offers is not an increase in the total extractable work beyond the initial ergotropy, but a physically transparent route to recovering work from a pure state that has become entangled with a pointer. It demonstrates how quantum correlations and conditional feedback can be used to preserve and recover work extraction capability when the battery is entangled with an unobserved pointer. Such insight is valuable for the design of quantum thermodynamic devices in which entanglement and measurement play essential roles.

\section{Physical Implementation}

The proposed protocol might be well-suited for experimental realization in several leading quantum technology platforms. The essential requirements are: (i) a two-level quantum battery with Hamiltonian $\hat{H} = \epsilon \ket{e}\bra{e}$, (ii) a continuous-variable pointer with position and momentum operators $\hat{X}$ and $\hat{P}$, and (iii) an interaction that generates the unitary $U = \exp(-i \lambda \hat{H} \otimes \hat{P}/\hbar)$, thereby entangling the battery's energy eigenstates with the pointer's position. While other platforms such as cavity QED or optomechanical systems could also realize this interaction, we focus here on the trapped-ion implementation, which offers a particularly direct route to the required conditional displacement.

In a trapped-ion system, the quantum battery can be realized by two long-lived internal electronic states of a single ion, denoted as $\ket{g}$ and $\ket{e}$. The pointer is realized by a collective motional mode, such as the center-of-mass (COM) mode, which behaves as a quantum harmonic oscillator with position and momentum operators $\hat{X}$ and $\hat{P}$.

The required state-dependent displacement can be generated by applying a bichromatic laser field with frequencies detuned from the electronic transition. This creates a spin-dependent optical dipole force. The interaction Hamiltonian can be written in the general form \cite{Bazavan2023, haljan2005spin}:
\begin{equation}
\hat{H}_{\text{int}} = \hbar \chi \, \hat{\sigma}_z \otimes \left( \hat{a} e^{-i\phi} + \hat{a}^\dagger e^{i\phi} \right),
\label{eq:trapped_ion_general}
\end{equation}
where $\hat{\sigma}_z = \ket{e}\bra{e} - \ket{g}\bra{g}$ is the Pauli-$Z$ operator, $\hat{a}$ and $\hat{a}^\dagger$ are the annihilation and creation operators of the motional mode, $\chi$ is the coupling strength (with units of frequency), and $\phi$ is a laser phase that can be controlled experimentally.

To obtain a coupling to the momentum operator $\hat{P}$, we choose $\phi = \pi/2$. Then Eq.~(\ref{eq:trapped_ion_general}) becomes
\begin{equation}
\hat{H}_{\text{int}} = \hbar \chi \, \hat{\sigma}_z \otimes i(\hat{a}^\dagger - \hat{a}).
\label{eq:trapped_ion_momentum}
\end{equation}
Using the standard relations for a harmonic oscillator,
$\hat{X} = x_{\text{zpf}}(\hat{a}^\dagger + \hat{a})$ and $\hat{P} = i\frac{\hbar}{2x_{\text{zpf}}}(\hat{a}^\dagger - \hat{a})$,
with $x_{\text{zpf}} = \sqrt{\hbar/(2m\omega)}$ the zero-point fluctuation amplitude, we find
$i(\hat{a}^\dagger - \hat{a}) = \frac{2x_{\text{zpf}}}{\hbar}\,\hat{P}$.
Thus,
\begin{equation}
\hat{H}_{\text{int}} = 2\chi x_{\text{zpf}} \, \hat{\sigma}_z \otimes \hat{P}.
\label{eq:trapped_ion_sigmaP}
\end{equation}
Since the battery Hamiltonian is $\hat{H} = \epsilon \ket{e}\bra{e} = \frac{\epsilon}{2}(I + \hat{\sigma}_z)$, we have $\hat{\sigma}_z = \frac{2}{\epsilon}\hat{H} - I$. Substituting this into Eq.~(\ref{eq:trapped_ion_sigmaP}) yields
\begin{equation}
\hat{H}_{\text{int}} = \frac{4\chi x_{\text{zpf}}}{\epsilon}\,\hat{H}\otimes\hat{P} - 2\chi x_{\text{zpf}}\, I\otimes\hat{P}.
\label{eq:trapped_ion_final}
\end{equation}
The second term is a global displacement of the pointer that is independent of the battery state. It can be eliminated by applying a suitable displacement operation to the pointer before the interaction or by using a spin-echo sequence that refocuses this global effect. The first term is exactly of the required form $\lambda' \hat{H}\otimes\hat{P}$ (with $\lambda' = 4\chi x_{\text{zpf}}/\epsilon$), which generates the desired conditional displacement. The unitary evolution for a time $t$ is therefore
\begin{equation}
U = \exp\!\left( -\frac{i}{\hbar}\int_0^t \hat{H}_{\text{int}}\,dt' \right)
= \exp\!\left( -\frac{i}{\hbar}\,\lambda\,\hat{H}\otimes\hat{P} \right),
\end{equation}
with the effective coupling parameter
\begin{equation}
\lambda = \frac{4\chi x_{\text{zpf}}\, t}{\epsilon}.
\label{eq:lambda_trapped_ion}
\end{equation}
The resulting conditional displacement of the pointer in position space is $d = \lambda \epsilon = 4\chi x_{\text{zpf}} t$, which has units of length.

The necessary experimental steps are:
\begin{enumerate}
    \item \textbf{Initialization:} The ion is prepared in the coherent state $\ket{\psi_S}$ using microwave or Raman pulses. The motional mode is prepared in its ground state, and then a displacement operation is applied to create the Gaussian state $\ket{\psi_P}$.
    \item \textbf{Interaction:} The state-dependent force described by Eq.~(\ref{eq:trapped_ion_momentum}) is applied for a time $t$, generating the conditional displacement unitary $U$.
 \item \textbf{Measurement:} A measurement of the ion's motional state is performed. This can be achieved by coupling the motion to the internal state via a sideband transition and then measuring the internal state via state-selective fluorescence \cite{meekhof1996generation}. This can be used to implement a measurement of the motional position with finite experimental resolution, providing an operational realization of the idealized position measurement considered in the theoretical model. In practice, finite resolution corresponds to a coarse-grained position measurement, which generally leaves the conditional battery state mixed and reduces the recovered ergotropy below the ideal value in Eq.~(\ref{eq:daemonic_ergotropy_result}); the ideal limit is approached only when the resolution is much smaller than the relevant pointer scales, such as the initial width \(\sigma\) and the displacement \(d\).
    \item \textbf{Feedback:} Based on the outcome $x$, a conditional unitary operation is applied to the battery. This can be implemented by a microwave or Raman pulse with phase and/or amplitude controlled by the measurement result. The speed of classical electronics in modern trapped-ion setups is sufficient for real-time feed-forward \cite{negnevitsky2018repeated}.
\end{enumerate}

Recent experiments have successfully demonstrated the charging of a quantum battery using a single-ion information engine \cite{Zhang2025}, and the conditional displacement operation is a standard tool in trapped-ion quantum information processing \cite{Leibfried2003}. This makes trapped ions a promising platform for an experimental realization of the proposed protocol.

The pointer acts as a continuous-variable quantum measurement ancilla whose position becomes correlated with the battery's energy. The projective measurement of the pointer's position provides information about the battery while partially preserving its coherence, a key feature of the daemonic ergotropy protocol. A detailed thermodynamic analysis of the measurement and feedback costs is crucial for a complete understanding of the protocol's performance; see, e.g., the recent review in Ref.~\cite{de2025friendly}. The work cost of the position measurement is associated with the energy required to couple the pointer to the detector and the dissipation incurred during the measurement process. This is consistent with the principles of information thermodynamics, which have recently been shown to hold universally in quantum feedback control \cite{minagawa2025universal}. Similarly, the feedback cost is the energy required to implement the conditional unitary, which in trapped-ion systems can be realized by laser pulses whose timing and phase are controlled by the measurement outcome. Recent theoretical advances have established fundamental limits for thermodynamic control with quantum feedback \cite{Kumasaki2025}, and our protocol provides a concrete framework for studying these limits in practice. Whether positive net work can be achieved depends on the measurement and feedback costs of the specific implementation and requires a detailed experimental thermodynamic analysis beyond the scope of the present work.

The implementation of the proposed protocol in these platforms is, therefore, not only feasible but also aligns with the state-of-the-art in quantum control and measurement, paving the way for experimental tests of daemonic ergotropy and the role of quantum coherence in work extraction.

\section{Conclusion}

We have proposed a protocol for harvesting work from a coherent quantum battery via a conditional spatial displacement, described by the unitary \(U = \exp(-i k t \, \hat{H} \otimes \hat{P}/\hbar)\). This operation entangles the battery with a continuous-variable pointer, thereby enabling a measurement-assisted feedback scheme that exploits quantum coherence. We examined the protocol for a two-level battery and a Gaussian pointer, deriving the conditional states after a position measurement and the resulting measurement-assisted extractable work \(\mathcal{W}_D^{(X)}\).

For a pure initial battery state, we found that \(\mathcal{W}_D^{(X)}\) coincides with the standard ergotropy at every interaction strength. Two features of the model are responsible: the interaction commutes with the battery Hamiltonian, leaving the energy populations intact, and the conditional states after the pointer measurement remain pure for all outcomes. As a result, the feedback protocol recovers the full initial ergotropy that would otherwise be lost when the pointer degrees of freedom are traced out. Although the scheme does not increase the extractable work beyond the initial ergotropy, it shows how quantum correlations and outcome-dependent operations can safeguard work extraction in the presence of an ancilla.

The protocol offers a physically transparent realization of a quantum Maxwell demon, with the pointer acting as a continuous-variable measurement ancilla whose position correlates with the battery energy. In contrast to a direct projective energy measurement, which erases coherence entirely, the pointer measurement retains conditional coherence for finite interaction strengths. Such coherence preservation could be advantageous when the conditional state is to be manipulated further or when coherence itself serves as a resource for quantum information tasks. This work thus adds to the growing literature on measurement, information, and coherence in quantum thermodynamics \cite{gumberidze2019measurement}.

The scheme is compatible with trapped-ion platforms. There, the conditional displacement can be generated by a spin-dependent optical dipole force, with the motional mode playing the role of the pointer. This implementation draws on established quantum control and measurement techniques, making the protocol accessible to current experiments.

Several extensions of this framework merit future study. A detailed accounting of measurement and feedback costs will be needed to determine the net work extractable under realistic conditions, where these operations consume finite energy. Extending the protocol to multi-level and continuous-variable batteries would broaden its scope and reveal effects tied to richer energy spectra. Alternative measurement strategies—such as generalized measurements or adaptive feedback—might recover or even surpass the initial ergotropy, especially for mixed initial states or when the interaction does not commute with the battery Hamiltonian. Optimizing the measurement to maximize daemonic ergotropy remains an open challenge \cite{Kua2026}. Extending the protocol to multi-cell batteries would also connect with recent 
work on the role of entanglement in charging power, where it has been shown 
that the relationship between entanglement and quantum advantage is subtle 
and not monotonic \cite{gyhm2024beneficial}.

This work establishes conditional spatial displacements as a versatile tool for measurement-assisted quantum thermodynamics. Although the specific model studied here does not achieve enhancement beyond the initial ergotropy, it provides a clear, analytically tractable example of how quantum correlations and conditional feedback can preserve and recover work extraction. By bridging quantum information and energy conversion, the protocol opens new avenues for investigating the thermodynamics of quantum measurement and feedback control.

\subsection*{Declaration of AI Usage}

During the preparation of this manuscript, the author utilized an AI-based language model solely for the purposes of language refinement, grammar correction, and assistance with LaTeX formatting. The AI was not used to generate, modify, or interpret any scientific content, derivations, results, or conclusions presented in this work. All theoretical developments, analytical derivations, numerical calculations, and scientific interpretations were carried out independently by the author. The author takes full responsibility for the accuracy and integrity of the content presented in this manuscript.

\end{document}